\documentclass[10pt,a4paper,aps,prl,twocolumn,showpacs,superscriptaddress]{revtex4-1} 
\usepackage[english]{babel}
\usepackage{url} 
\usepackage{hyperref}
\usepackage{amsmath}
\usepackage{slashed} 
\usepackage{graphicx}

\usepackage{color}
\usepackage{xcolor}
\usepackage{bm} 
\usepackage{cleveref}
\usepackage[normalem]{ulem}
\usepackage{units}

\usepackage{multirow}
\usepackage{tabularx}

\usepackage{subcaption}

\begin{document}
\allowdisplaybreaks

\title{Precision Measurement of Trident Production in Strong Electromagnetic Fields}
\author{Christian F. Nielsen}
\affiliation{Department of Physics and Astronomy, Aarhus University, 8000 Aarhus, Denmark}
\author{Robert Holtzapple}
\affiliation{Department of Physics, California Polytechnic State University, San Luis Obispo, California 93407, USA}
\author{Mads M. Lund}
\affiliation{Department of Physics and Astronomy, Aarhus University, 8000 Aarhus, Denmark}
\author{Jeppe H. Surrow}
\affiliation{Department of Physics and Astronomy, Aarhus University, 8000 Aarhus, Denmark}
\author{Allan H. S\o rensen}
\affiliation{Department of Physics and Astronomy, Aarhus University, 8000 Aarhus, Denmark}
\author{Marc B. Sørensen}
\affiliation{Department of Physics and Astronomy, Aarhus University, 8000 Aarhus, Denmark}
\author{Ulrik I. Uggerh\o j}
\affiliation{Department of Physics and Astronomy, Aarhus University, 8000 Aarhus, Denmark}
\collaboration{CERN NA63}

\date{\today}

\begin{abstract} 
We demonstrate experimentally that the trident process $e^-\rightarrow e^-e^+e^-$ in a strong external field, with a spatial extension comparable to the effective radiation length, is well understood theoretically. The experiment, conducted at CERN, probes values for the strong field parameter $\chi$ up to 2.4. Experimental data and theoretical expectations using the Local Constant Field Approximation show remarkable agreement over almost 3 orders of magnitude in yield.

\end{abstract}

%\pacs{41.60.-m,61.85.+p}
\maketitle
\section{Introduction}

When an electron impinges on an electrostatic potential barrier, it may penetrate or be reflected. Classically, for electron energies less than the barrier height, the electron is always reflected. In non-relativistic quantum mechanics, an exponentially damped tunneling into the barrier is predicted{, with no transmission when the potential remains higher than the electron energy beyond the classical turning point. In relativistic quantum theory, however, an undamped electron-current is present beyond the classical turning point provided the barrier rises sufficiently abruptly and high, even if it is infinitely high. This was shown in 1929 by Oscar Klein \cite{Klei29} for a step barrier in one of the first applications of the Dirac equation. 
It became known as the 'Klein paradox'. As later shown by Fritz Sauter \cite{Saut31a,Saut31b}, inspired by a supposition by Niels Bohr, the potential has to rise with the rest energy of the electron, $mc^2$, over its Compton wavelength, $\hbar/mc$, for transmission 
to occur with substantial probability. The corresponding field strength,
\begin{equation}
\mathcal{E}_0 = m^2c^3/e\hbar \simeq1.32\times10^{16}~\mathrm{V/cm},
\end{equation}
later became known as the critical or Schwinger field.

Previous studies of the Klein paradox have been limited to theory \cite{Grei85,Krek04,Giac08}, e.g.\ with phenomena analogous to the Klein paradox, possibly observable in graphene \cite{Kats06,Calo06,Buch06,Boggild2017,Nguyen2018}. Some
studies have been partly motivated by heuristic arguments linking the Klein paradox, strong field pair production and Hawking radiation from black holes \cite{Mull77,Hols98,Hols99}. Today, the Klein paradox is explained by the creation of electron-positron pairs at the boundary.

In the following, we demonstrate experimentally that the trident process $e^-\rightarrow e^-e^+e^-$ becomes significant when a strong external field is 'turned on'. In our experiment this happens when a crystal is rotated, relative to the beam direction, from a 'random' orientation, where the field 
experienced by a penetrating particle is the sum of essentially randomly placed screened Coulomb fields, to an axial orientation, where the 
atoms along the axis act coherently in deflecting the particle. This effectively creates a continuous field which is macroscopic along the main direction of motion and of critical magnitude in the particle rest frame. 
Regardless of the orientation of the crystal, there are two contributions to the trident process, the direct and a "two-step" or "cascade" contribution, where the incoming electron first emits a real photon which then converts to an electron-positron pair. In our case, the direct trident production, the emission of real photons, and conversion of photons into pairs, all proceed in essentially constant fields.

The trident process for electrons penetrating amorphous material where they interact with screened target nuclei, has been investigated in detail since the thirties, see e.g.\ \cite{Crane1939,Powell1949,Occhialini1949,Hooper1951,Barkas1952,Hooper1952Trident,Block1954,Heitler1954,Kelner1967,Grossete1968,BOHM1973,Baier2008}.
As a result, the amorphous yield of trident events is considered well known and the contributions to the total trident yield for the direct and two-step processes are similar for material thicknesses of a few percent of the radiation length $X_0$ \cite{Kelner1967,Baier2008}.

For %pure - hvad betyder "pure"? (brugt som modsætning til atomer/kerner)
extended electromagnetic fields, the trident process depends on the strong-field parameter $\chi$ defined through \cite{ritus_1985,Baie98,Berestetskii_b_1989}
\begin{equation}
             \chi^2=(F_{\mu\nu}{p}^\nu)^2/{m^2c^2}\mathcal{E}_0^2,
 \label{eq:chi}
 \end{equation}
 where  $F_{\mu\nu}$ is the electromagnetic field strength tensor and ${p}^\nu$ is the four-momentum. 
For values of $\chi$ smaller than unity, pair production by photons is exponentially suppressed \cite{ritus_1985,Baie98}, therefore the trident process is also suppressed. Consequently, the scale of the electric fields required for the trident process to occur are of the order $\mathcal{E}_0$. Such fields are not yet possible to produce in the laboratory. 
Nevertheless, a relativistic particle, with a relatively large Lorentz factor e.g. $\gamma \simeq 10^5-10^6$, will experience an electric field in its own rest frame that, with a suitable field configuration, may be Lorentz boosted and can reach $\chi > 1$, which enables the trident process.
The trident process in plane wave background fields has been, and is being investigated in great detail, both experimentally (SLAC-E144 experiment \cite{SlacE144}) and theoretically  (see e.g.\ the comprehensive 
review \cite{FedotovReview2022} and references therein). Most theoretical investigations are driven by several upcoming and ongoing experiments colliding GeV electron beams with tightly focused laser pulses to probe the trident process \cite{Luxe,E320Talk}. 

Aligned crystals have been a known source of strong electric fields for decades, see e.g. \cite{Ugge05,Baie98} for details on strong field effects in crystals. Depending on the element and orientation of the crystal, fields of the order $10^9-10^{11}$ V/cm are readily available in the laboratory frame. Due to the coherent action of the screened nuclei along crystallographic directions, this field is of truly macroscopic extent for Ge and Si crystals, where single crystals can be grown in principle to meters in length. For a charged particle moving in a field that is purely electric in the laboratory, and essentially transverse to the direction of motion as in the case of an aligned crystal, the strong field parameter reduces to $\chi \simeq \gamma \mathcal{E}/\mathcal{E}_0$. At CERN, electrons and positrons with up to $\gamma \simeq 10^6$ are available making it possible to achieve  values of $\chi > 1$ in the rest frame of the electron in an aligned crystal.

In 2007 an attempt to measure the production of electron-positron pairs through the trident process $e^-\rightarrow e^-e^+e^-$ in strong electromagnetic fields was made with participation from one of the present authors (UIU) \cite{Ulrik2009Trident}, that resulted in significant discrepancies  theory. 
In this paper we report on a recent precision measurement on the trident process, impinging 200 GeV electrons on the same 400 $\mu$m germanium crystal target as in 2007, oriented along the $\langle110\rangle$ axis, showing an extraordinary agreement between theory and experiment. Thus, the discrepancies observed in the 2007 experiment are yet to be explained.  It is our conjecture that the 2007 experimental values are incorrect possibly due to improper alignment of the crystal.

\begin{figure*}[ht!]
	\begin{center}
		\includegraphics[width=0.85\linewidth]{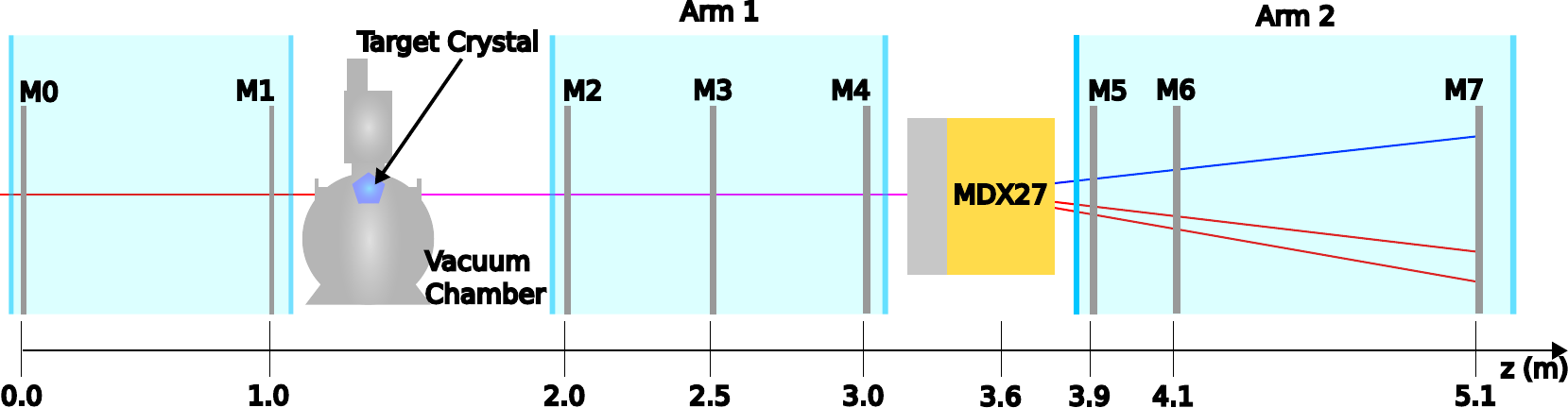}
\caption{Experimental setup. A schematic representation of the experimental setup in the H4 beam line in the SPS North Area at CERN. The symbols ``Mi'', with $i=0,\ldots,7$, denote `Mimosa-26' position sensitive detectors.}
		\label{fig:Setup}
	\end{center}
\end{figure*}

 \section{Experiment}
 The experiment was performed by the NA63 collaboration at the H4 beamline of the CERN SPS that provided a 200 GeV electron beam having a $\sigma_x \simeq \sigma_y \simeq 105$ $\mu$rad divergence impinging on the %an oriented 
 400 $\mu$m thick $\langle110\rangle$ oriented germanium single crystal.  \Cref{fig:Setup} is a schematic of the setup where M0-M7 are MIMOSA-26 position sensitive CMOS-based pixel detectors \cite{Mimosa26}.  The detectors have a resolution of a few $\mu$m and an active area of $1.1\times 2.1$ cm$^2$ containing $576\times1152$ pixels. The crystal target is mounted on a goniometer that allows us to set the crystal orientation with $\mu$rad precision. The MDX27 magnet provides an integrated magnetic field of $0.072$ Tm and, together with the detectors in Arm 1 and Arm 2, forms a magnetic spectrometer allowing us to measure the energy of each charged particle from the deflection angle in the magnet. The crystal is situated inside a vacuum chamber at $\simeq 300$ K. To reduce scattering and background, all mimosas are placed in closed compartments that are continuously flushed with helium. 
  The total material contributing to the background before the MDX27 magnet, in units of the radiation length, amounts to $\simeq 1.1\%$.

 The incoming electron rate was set to ensure that only a single primary electron was present in the setup per event. We therefore have essentially complete information about the particle and its secondaries.

Three measurement series were performed; a background measurement with no target, a 'random' measurement where the crystal is rotated far away from any major crystallographic orientation thus acting as an amorphous target, and a measurement with the aligned crystal. In the latter, the crystal was aligned with the beam centroid along the $\langle110\rangle$ axis through a careful procedure where enhancement of radiation upon the passage of crystalline planes creates a stereogram from which the location of the axis can be found with a precision of about $20~\mu$rad.

\begin{figure*}[ht!]
\centering
\begin{subfigure}{.5\textwidth}
		\includegraphics[width=\linewidth]{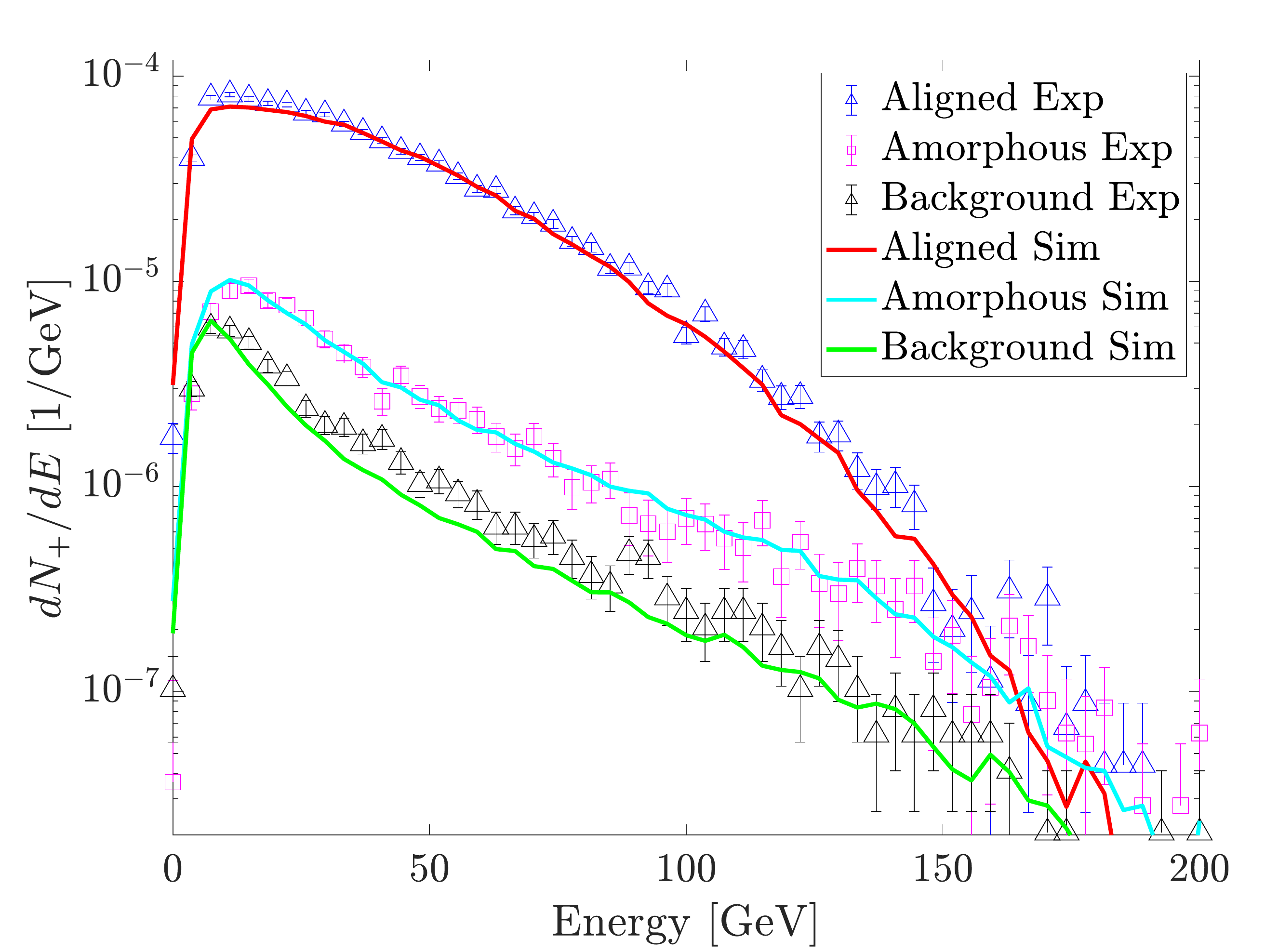} 
\end{subfigure}%
\begin{subfigure}{.5\textwidth}
		\includegraphics[width=\linewidth]{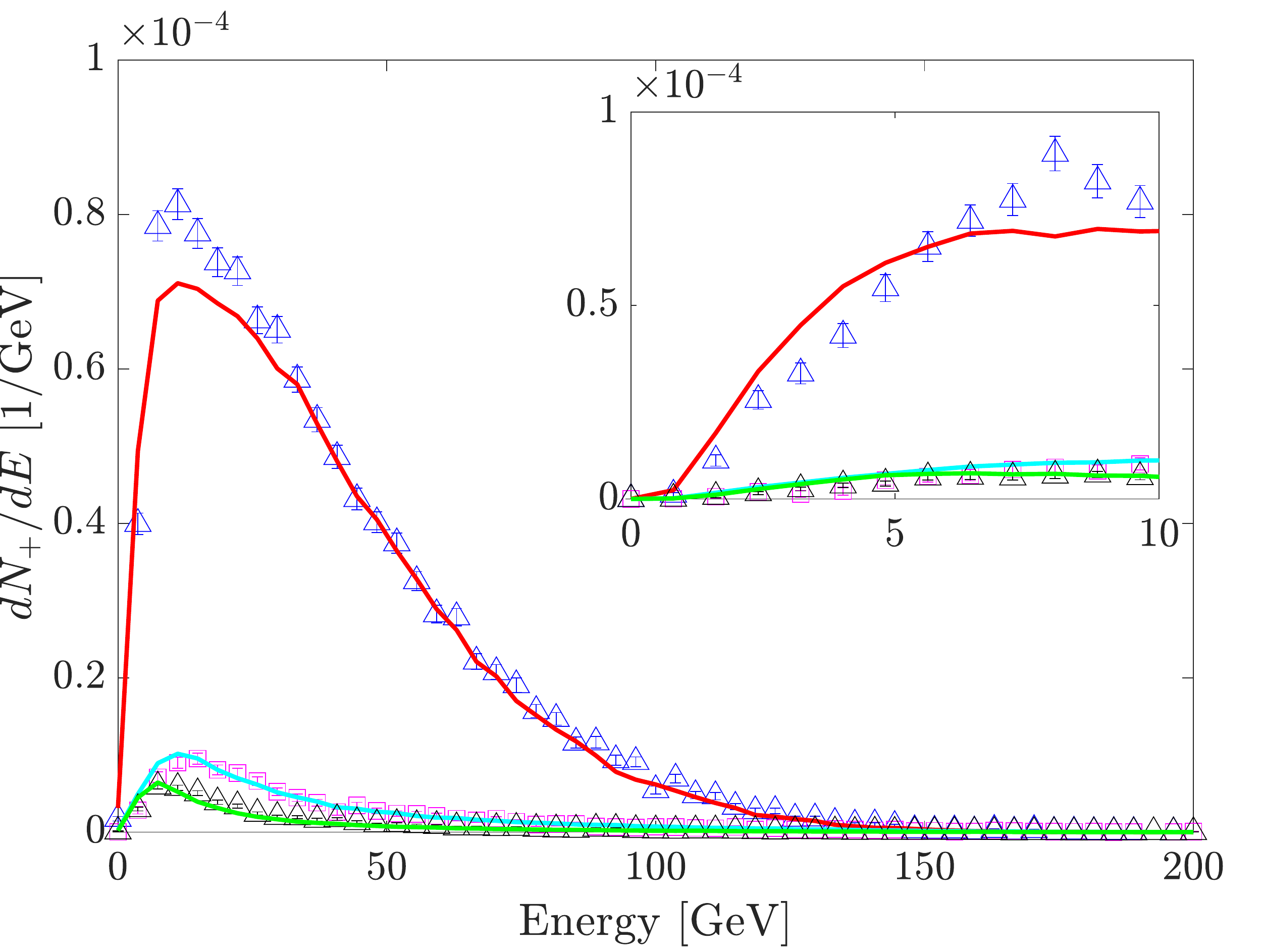} 
\end{subfigure}
\caption{Positron spectrum of reconstructed trident events. Solid lines are simulations while squares and triangles are experimental data points. Blue and Red show the trident spectrum in aligned orientation, magenta and cyan show the amorphous yield while black and green is the background. The right figure is the same as the left but with a linear %y-axis. 
vertical axis. The inset shows a zoom of the low energy part, which largely overlaps with the energy range measured in 2007~\cite{Ulrik2009Trident}.}
\label{fig:Spectrum}
\end{figure*}

\section{Data Analysis and Theoretical Comparison}
The data analysis routine begins by finding complete single particle tracks after the target by matching tracks found in Arm 1 and Arm 2 in the center of the magnet. A track in Arm 1 is only accepted if it simultaneously overlaps with a hit in M0 and M1. To find possible trident events, all complete single particle tracks are combined and matched through strict matching criteria.  The energy resolution of the magnetic spectrometer for a single particle track was measured to be $\sigma_E/E \simeq 6.7 \%$ at 200 GeV (including $dp/p \simeq 1\%$ from the beamline) and it remains relatively uniform down to particle energies of a few GeV.  

To find possible trident events, all complete single particle tracks are combined and matched through strict matching criteria.  The energy resolution of the magnetic spectrometer for a single particle track was measured to be $\sigma_E/E \simeq 6.7 \%$ at 200 GeV (including $dp/p \simeq 1\%$ from the beamline) and it remains relatively uniform down to particle energies of a few GeV.  

\Cref{fig:Spectrum} shows the experimentally measured positron spectrum of all three measurement series together with theoretical predictions. The theoretical curves labeled "Sim" result from a full scale Monte-Carlo simulation of the experimental setup, starting with a beam of electrons identical to what was provided by the SPS. 
The simulation propagates primary electrons individually through each element of the setup, where they are subject to multiple Coulomb scattering and emission of photons, that can later pair produce, and direct trident production. When a particle penetrates a detector, its position is saved in a data file identical to the experimental data file containing $(x, y, z)$ positions from each detector from particle hits. The simulated data files are then analyzed with the experimental analysis routine. As in the experiment, we simulate all three measurement conditions  and the background is subtracted from the amorphous and aligned curves after the analysis. To account for detector efficiencies, we fit a linear energy dependent efficiency, $f(E) = aE+b$, to the ratio between the experimental and simulated curves for the amorphous case after background subtraction. All simulated curves shown in \cref{fig:Spectrum} are then the result of the experimental analysis routine multiplied by the linear energy efficiency factor with the fitting parameters  
$
    a = -0.0012 \pm 0.001 \text{GeV}^{-1} \quad \text{and} \quad b = 0.55 \pm 0.08$. 
The value for $b$, found by this procedure, agrees well with expectations based on the efficiency of the Mimosa detectors that depend on set threshold values, and the value of $a$ is small.

When particles penetrate amorphous material, the probability of the two-step trident process is modeled by Bethe-Heitler photon emission followed by Bethe-Heitler pair production as described in e.g, \cite{PDG_2022}. The scattering centers are completely decoupled and non-linear effects, e.g.\ photons emitted in the background that pair produce in the crystal or vice versa, are then taken into account by the simulation of the experiment. Therefore, it is important that all elements are included correctly in the setup and that the background simulation matches the measurement. The direct trident process is modeled using the theory described in \cite{Kelner1967,Baier2008}.
Amorphous material in the background and the crystal oriented in 'random' are modeled identically, the processes depend on the nuclear charge $Ze$ and the radiation length $X_0$.
\Cref{fig:Spectrum} shows that the theoretical curves and experimental data are in good agreement for the amorphous case, as in the 2007 experiment \cite{Ulrik2009Trident}. Using the detector-efficiency correction on the background, we also see good agreement, which strongly indicates that the background contribution is well understood. 

When a particle hits the aligned crystal, a separate simulation code is employed which is described in detail in \cite{CFN2019GPU,NIELSEN2022}. The simulation solves the equation of motion for each time step based on the Lorentz force in the electric fields modeled by the Doyle-Turner continuum string potentials \cite{Doyl68,JUAnotes}. For every time step the local value of $\chi$ is evaluated and the probability of photon emission, based on the Local Constant Field Approximation (LCFA), is evaluated. This photon is then propagated forward through the crystal where the probability of pair production, using the LCFA, is again evaluated and the produced pair is propagated through the setup (details on the implementation of these processes can be found in \cite{NIELSEN2022}). 

The applicability of the LCFA requires that the angular excursions of the projectile over a formation length are wider than the light cone $1/\gamma = 2.6$ $\mu$rad. In the channeling regime, the angular excursions of a charged particle are on the order of the critical Lindhard angle \cite{Lind65,JUAnotes}, which in our case is  57 $ \mu$rad. For entry angles larger than the critical Lindhard angle, but smaller than the Baier angle $U_0 / mc^2 = 0.4 $ mrad, the angular deflections remain larger than $1/\gamma$ \cite{SORENSEN19962}. Here $U_0 = 215$ eV is the continuum string potential depth, in this case for a single $<110>$ row of Ge atoms. As a result, with a beam divergence of $\simeq 105$ $\mu$rad, the LCFA is appropriate for nearly all particles within the beam.
As a  measure of the applicability of the constant-field approximation under channeling conditions, the authors of \cite{Baie98} introduced the parameter $\rho_c = \xi^2 = 2U_0\gamma/mc^2 $  (where $\xi$ is known as the classical non-linearity parameter \cite{AntoninoReview,FedotovReview2022} in the strong-field laser community), which under channeling conditions has a value of $\simeq 330$ for 200 GeV electrons. 
The large value of $\rho_c$  verifies that treating the local field as constant is a good approximation.

For the aligned crystal, the direct trident process is modeled through the Weizs\"{a}cker-Williams method of virtual quanta \cite{Jackson_b_1975,Baie98}, together with the LCFA for the pair production vertex. For constant electric fields, the Weizs\"{a}cker-Williams method has been investigated and deviates around $10\%$ from methods evaluating the actual direct two-vertex Feynman diagrams \cite{KingRuhl2013}. This difference has marginal influence in our case due to the dominance of the two-step process.

\Cref{fig:Spectrum} shows that the theoretical prediction for the aligned case agrees remarkably well with the experimental data points throughout the entire spectrum which spans several orders of magnitude.

The relative importance of the direct process and the two-step process can be estimated by comparing the virtual Weizs\"{a}cker-Williams photon intensity (radiated energy per photon-energy interval) with the real photon intensity. The virtual photon intensity is given by the fine-structure constant up to a logarithmic factor. The real photons have a fairly flat intensity spectrum given by $L/X$, where $L$ is the target thickness and $X$ the effective radiation length defined as $X=E/(dE/dx)$ where $dE/dx$ is the energy-loss rate per unit length due to radiation. Therefore, the two processes are comparable in strength for a target thickness of order a percent of the effective radiation length (see also \cite{KingRuhl2013,BenKing2018}). This is the case for the amorphous setting in our experiment ('random' setting), where $X=X_0=2.30$ cm and $L/X_0 =1.7$ \%. Accordingly, the simulations show that $\sim50\%$ of all tridents come from the direct process in the amorphous setting. For the aligned case, the effective radiation length $X$ is much shorter than $X_0$ due to the strong-field effects. The stronger radiation causes the direct process to only contribute a few percent to the total pair rate.

In the two-step process, the theoretical model applied averages over the photon polarization. The polarization has been predicted to have significant effects in strong constant fields \cite{BenKing2013Polarization,Sepit2020Polarization}. For axially aligned crystals, the experimentally measured photon spectrum is polarization averaged due to each projectile having a unique trajectory through the crystal. If a real photon is emitted with a specific polarization, this photon will also follow a unique trajectory due to the uniqueness of the emitting particle.  This means the pair production process also becomes polarization averaged, and modeling the real photon as unpolarized, is a good approximation. 

\section{Conclusion}
By impinging 200 GeV electrons on a $\langle 110\rangle$ oriented 400 $\mu$m thick germanium crystal, we investigated the trident process in strong electric fields.   Our experimental results are in remarkably good agreement with theory based on LCFA, the locally constant field approximation.

The inset in \cref{fig:Spectrum} shows the low-energy tail (which is subject to a major non-linear drop in detection efficiency) corresponding to the energy range studied in the 2007 experiment \cite{Ulrik2009Trident}. Here too, there is good agreement between theory and experiment. Accordingly, the discrepancy in 2007 most likely was due to an unknown experimental error.

In our case the experimental spectrum is dominated by the two-step process. 
To quantify the direct process using an aligned crystal, the crystal will have to be much thinner than what was used here. An ultra thin crystal greatly reduces the production rate, making it a challenging measurement, even if the crystal is cooled.

\section{Acknowledgments}
The numerical results presented in this work were partly obtained at the Centre for Scientific Computing Aarhus (CSCAA) and with support from Nvidia's GPU grant program. This work was partially supported by the U.S. National Science Foundation (Grant No. PHY-1535696, and PHY-2012549) and from the Danish National Instrument Center for CERN Experiments (NICE).

%\bibliography{references}

\end{document}